\documentclass[conference]{IEEEtran}

\usepackage{graphicx,cite}
\usepackage{multirow}
\usepackage{epsfig}
\usepackage[cmex10]{amsmath}   
\usepackage{array}
\usepackage{subfig}
\usepackage{algorithm}
\usepackage{algorithmic}
\usepackage{amssymb}
\usepackage{array}
\usepackage{fixltx2e}
\usepackage{amsfonts}

\begin{document}
\title{Power Optimal Non-contiguous Spectrum Access in Multi Front End Radio Enabled Point-to-Point Link}


\author{\IEEEauthorblockN{Muhammad Nazmul Islam, Narayan B. Mandayam, Ivan Seskar}
\IEEEauthorblockA{Department of Electrical \& Computer Engineering\\
WINLAB, Rutgers University, USA \\
Email: \{mnislam,narayan,seskar\}@winlab.rutgers.edu}
\and
\IEEEauthorblockN{Sastry Kompella}
\IEEEauthorblockA{Information Technology Division, \\
Naval Research Laboratory, USA\\
Email: sk@ieee.org}}

\maketitle

\begin{abstract}

Non-contiguous spectrum chunks allow wireless links to flexibly access a wide amount of bandwidth.
Multi-Channel Multi-Radio (MC-MR) and Non-Contiguous Orthogonal Frequency Division Multiplexing (NC-OFDM)
are the two commercially viable strategies to access non-contiguous spectrum chunks.
MC-MR accesses multiple non-contiguous chunks by activating multiple front ends 
which, in turn, increases the circuit power consumption of each of the activated front ends.
NC-OFDM accesses non-contiguous spectrum chunks with a single front end by nulling remaining subchannels
but increases spectrum span which, in turn, increases the
power consumption of ADC and DAC. This work focuses on a point-to-point link where transmitter and
receiver have multiple front ends and can employ NC-OFDM technology. 
We investigate optimal spectrum fragmentation in each front end from a system power -- summation of transmit power and circuit power -- perspective.
We formulate a mixed integer non-linear program (MINLP) to perform power control and scheduling, and
minimize system power by providing a greedy algorithm $(O(M^3 I))$ 
where $M$ and $I$ denote the number of channels and radio front ends respectively.

\end{abstract}


\section{Introduction}   \label{sec:Intro}

Uncoordinated interference in unlicensed networks  result in an arbitrary number of non-contiguous spectrum chunks available for transmission, 
with some of these chunks separated by wide bands of interference limited spectrum. 
In TV white space~\cite{802.22}, available channels themselves are non-contiguous in nature. 
Accessing only the ``best" channel~\cite{802.22Channel} or the ``best" contiguous portion of these channels~\cite{WhiteFi1} leads to an underuse of the available spectrum.
Hence, non-contiguous spectrum access is becoming increasingly popular among software defined radio researchers~\cite{UCSB2,UCSB3}.

Multi-Channel Multi-Radio (MC-MR)~\cite{MC-MR1} and Non-Contiguous Orthogonal Frequency Division Multiplexing (NC-OFDM) are the two
commercially viable choices to access non-contiguous spectrum chunks. MC-MR access multiple non-contiguous spectrum
chunks by activating multiple radio front ends where each front end captures a contiguous portion of available spectrum
and, hence, activates the circuit power consumption of each front end. 
NC-OFDM accesses non-contiguous spectrum chunks with a single front end radio by nulling intermediate spectrum
and, hence, increases spectrum span~\cite{Nazmul_NCOFDM1} which, in turn,
increases the power consumption of ADC and DAC.

\begin{figure}[t]
\centering
\includegraphics[scale=0.3]{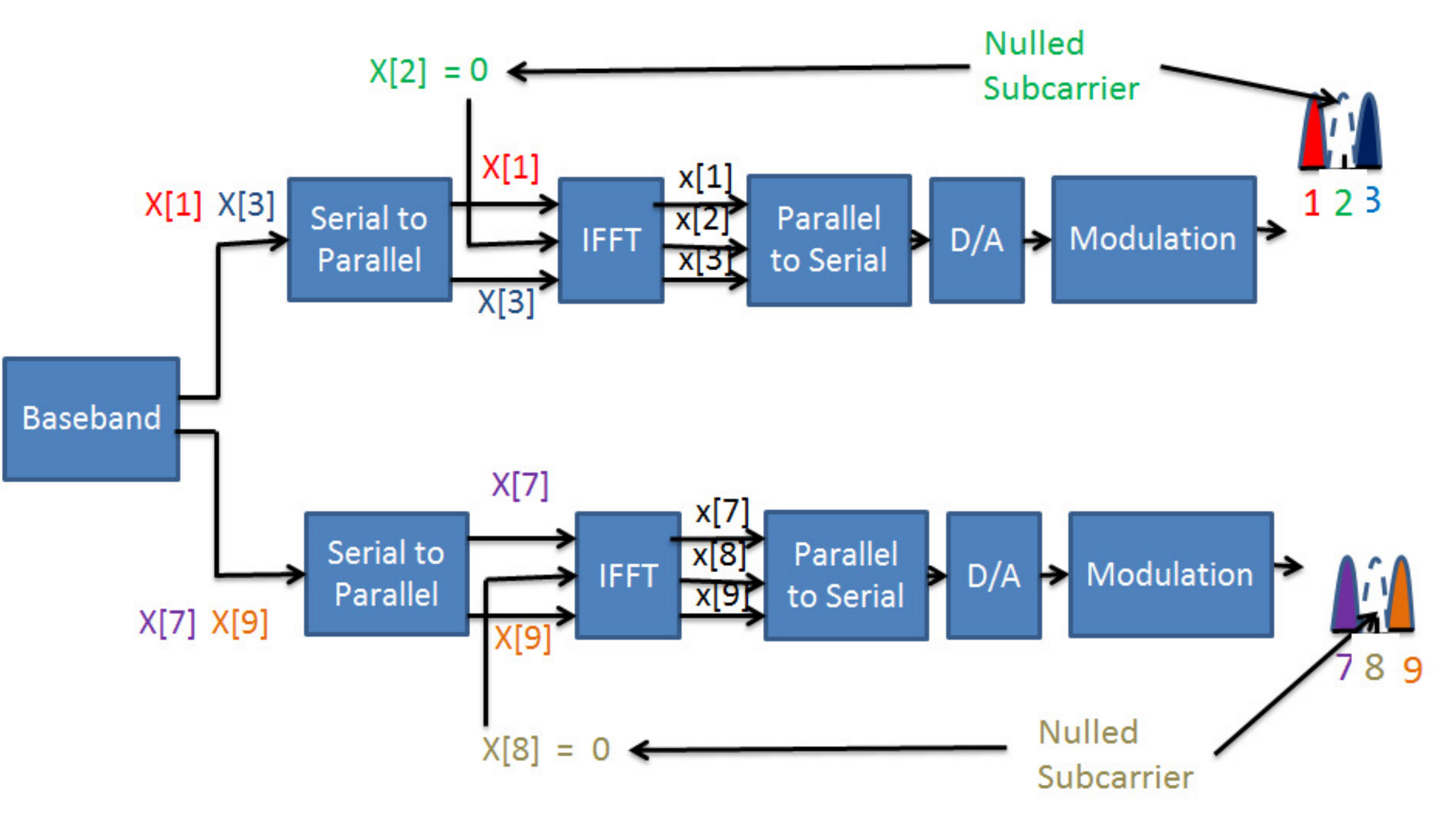}
\caption{Multi-Channel Multi-Radio based Non-Contiguous Orthogonal Frequency Division Multiplexing operation}
\label{fig:MCMR_NCOFDMA}
\end{figure}

Fig.~\ref{fig:MCMR_NCOFDMA} shows the advantages and pitfalls of MC-MR and NC-OFDM. Here, channel $1$, $3$, $7$ and $9$ are available.
One radio front end (shown at the top) accesses channel $1$ \& $3$ and nulls channel $2$ using NC-OFDM.
Other radio front end (shown at the bottom) accesses channel $7$ \& $9$ and nulls channel $8$ using NC-OFDM.
Due to the use of two front ends, the circuit power consumption of different components inside modulation block - e.g. filters, mixers, etc. --
increases by a factor of $2$. Due to the nulling of intermediate spectrum, each front end spans three channels (instead of two) and
the circuit power consumption of ADC and DAC increase by a factor of $1.5$.

This work performs optimal power control and scheduling to minimize the system power -- summation of transmit and circuit power --
of a point-to-point link where both nodes are equipped with MC-MR and NC-OFDM technology.
Given an available list of non-contiguous spectrum chunks, we find the optimal spectrum fragmentation in each radio front end
using a greedy algorithm $(O(M^3 I))$ where $M$ and $I$ denote the number of channels and radio front ends respectively.

The authors of~\cite{Goldsmith,Sahai,CktPower2} investigated system power consumption in
radio front ends but focused on contiguous spectrum access.
Our other work~\cite{Nazmul_NCOFDM1} investigated system power minimization in non-contiguous spectrum access
but focuses on single front end radio enabled nodes.
This work minimizes system power of multi-front-end radio enabled nodes during non-contiguous
spectrum access.

\section{System Model}  \label{sec:Model}

We focus on a point-to-point link where both transmitter and receiver are equipped with a set
of front ends $\mathcal{I}$. Nodes access a set of available channels $\mathcal{M}$
to meet demand $r$. Let, $M = |\mathcal{M}|$ and
$I = |\mathcal{I}|$.
Let $g^m$ denote the link gain at channel $m \in \mathcal{M}$. Let $p_i^m$ and $x_i^m$ represent the power control
and scheduling decision variables in channel $m \in \mathcal{M}$ at front end $i \in \mathcal{I}$.

We assume that baseband signal processing techniques like multiuser detection
and iterative decoding are not employed. Baseband power consumption
is negligible in this scenario~\cite{Goldsmith}. We focus on system power minimization during active mode (when the link is transmitting data)~\cite{Goldsmith}.
The authors of~\cite{Goldsmith} show the following system power consumption model of transmitter ($p_t$)
and receiver ($p_r$):
\begin{equation}
p_t = \alpha_1 + \alpha_2 f_s + k_{pa} p \, , \, \, p_r = \beta_1 + \beta_2 f_s \, \label{eq:GoldsmithModel}
\end{equation}
Here, $p$ is the emitted power at radio frequency, $k_{pa} = \frac{PAPR}{\eta}$,
$k_{pa} p$ is the power consumption of programmable amplifier, $PAPR$ is the peak-to-average-power-ratio of
the modulation scheme, $\eta$ is the efficiency of the programmable amplifier and
$f_s$ is sampling rate. 
$\alpha_2$ and $\beta_2$ are the slopes of DAC and ADC's power consumption versus sampling rate curves.
$\alpha_1$ ($\beta_1$) is the power consumption of all blocks of transmitter (receiver) - e.g. filter, mixer, etc. -
excluding DAC (ADC) and programmable amplifier~\cite{Goldsmith}. 

The total power consumption of a transmitter and receiver equipped with $\mathcal{I}$ front ends and $\mathcal{M}$ 
available channels can be written as:
\begin{equation}
\sum_{i \in \mathcal{I}} \bigl( k_{pa} \sum_{m \in \mathcal{M}} p_i^m + \alpha_{1,i} + \alpha_2 f_{s,i} + \beta_{1,i} + \beta_2 {f_{s,i}} \bigr)
\end{equation}
where $\alpha_{1,i}$ ($\beta_{1,i}$) is the power consumption of all blocks of transmitter's (receiver's) $\mathrm{i}$-th front end
excluding DAC (ADC) and programmable amplifier. 
The sampling rate $f_{s,i}$ of $\mathrm{i}$-th front end depends on its spectrum span, which in turn is determined by the choice of channels (subcarriers) 
selected for its intended transmission.  
\emph{Spectrum span is defined as the gap between the furthest edges 
of the used channels}. 
Let $q_i$ denote the spectrum span of the transmitter's
and receiver's $\mathrm{i}$-th front end. Using the analysis of~\cite{Nazmul_NCOFDM1},
\begin{align*}
q_i & = && W \cdot \max \bigl(\bigl(\max_{m \in \mathcal{M}} \bigl(m \cdot x_i^m \bigr)  
-   \nonumber \\ 
& && \min_{m \in \mathcal{M}} \bigl(m \cdot x_i^m + |M| \cdot (1 - x_i^m) \bigr) + 1 \bigr), 0 \bigr)
\label{eq:SpanTx}
\end{align*}
Sampling rate should be at least twice the amount of spectrum span. We assume, $f_{s,i} = 2 q_i$.
Our overall optimization problem is shown below:

\emph{Problem I}

\begin{subequations}

\begin{equation}
\min \sum_{i \in \mathcal{I}} \bigl( k_{pa} \sum_{m \in \mathcal{M}} p_i^m + \alpha_{1,i} + 2 \alpha_2 q_i + \beta_{1,i} + 2 \beta_2 q_i \bigr)
\label{eq:Objective}
\end{equation}
s.t.
\begin{align}
q_i & \geq && W \cdot \max \bigl(\bigl(\max_{m \in \mathcal{M}} \bigl(m \cdot x_i^m \bigr)  
-   \nonumber \\ 
& && \min_{m \in \mathcal{M}} \bigl(m \cdot x_i^m + |M| \cdot (1 - x_i^m) \bigr) + 1 \bigr), 0 \bigr)
\label{eq:ScheduleConstraintMain}
\end{align}

\begin{equation}
\alpha_{1,i}  \geq   \alpha_{1,i} x_i^m \, , \, 
\beta_{1,i}  \geq   \beta_{1,i} x_i^m \, \,
 \forall m \in \mathcal{M} \, \forall i \in \mathcal{I} 
\label{eq:AnalogPower}
\end{equation}

\begin{equation}
\sum_{m \in \mathcal{M}} \sum_{i \in \mathcal{I}} W \log_2 \bigl( 1 + \frac{p_i^m g^m}{N_0 W} \bigr) \geq r
\label{eq:PowerConstraintMain}
\end{equation}

\begin{equation}
p_i^m \leq A x_i^m \, \forall m \in \mathcal{M} \, , \, \forall i \in \mathcal{I}
\label{eq:CouplingConstraintMain}
\end{equation}

\begin{equation}
\sum_{i \in \mathcal{I}} x_i^m \leq 1 \, \forall m \in \mathcal{M} \label{eq:FrontEndConstraint}
\end{equation}

\begin{equation}
x_{ij}^m \in \{0,1\} \, , \, p_{i}^m \geq 0 \, ,
q_i \geq 0 \, , \, \alpha_{1,i} \geq 0 \, , \, \beta_{1,i} \geq 0 \,
\forall i \in \mathcal{I}, \, \forall m \in \mathcal{M} \, ,
\label{eq:Variables1} 
\end{equation}

\label{eq:Variables2}

\end{subequations}

Here, Eq.~\eqref{eq:AnalogPower} denote that 
blocks like filter, mixer, etc. of $\mathrm{i}$-th front end consume power only if the node activates $\mathrm{i}$-th front end.
Eq.~\eqref{eq:PowerConstraintMain} denotes that summation of capacities across different channels and front ends must exceed demand $r$
where $N_0$ is noise spectral density.
Eq.~\eqref{eq:CouplingConstraintMain} couples power control and scheduling variables using a pre-defined big number $A$.
Eq.~\eqref{eq:FrontEndConstraint} represents that each channel can only be used by one of the front ends.


\section{Solution Methodology}  \label{sec:Insights}

\subsection{Theoretical Insights}  \label{sec:Theory}

The objective of problem I is linear. The constraints are convex.
$M I$ scheduling decision variables
are binary. Problem I is a mixed integer convex program and
can be solved optimally by solving $2^{MI}$ convex optimization programs.
This section designs a low complexity algorithm by analyzing the characteristics of problem I.

Problem I is in essence the combination of two separate 
optimization problems. The objective is to minimize
the summation of transmit and circuit power.
Eq.~\ref{eq:PowerConstraintMain} shows the constraints associated with transmit power.
Eq.~\eqref{eq:ScheduleConstraintMain} and ~\eqref{eq:FrontEndConstraint} denote the constraints
circuit power. Eq.~\ref{eq:CouplingConstraintMain} couples power and scheduling variables.
Hence, depending on the values of $\alpha_1$, $\alpha_2$, $\beta_1$ and $\beta_2$, problem I can have three sub-cases.

\subsubsection{Case I: Contiguous Spectrum Access}

When $\alpha_1$, $\alpha_2$, $\beta_1$ and $\beta_2$ are very large, circuit power completely dominates system power. 
Ignoring the contribution of transmit power $k_pa \sum_{m \in \mathcal{M}} p_m$
from Eq.~\eqref{eq:Objective}, problem I gets reduced to the following:

%

\begin{equation}
\min \sum_{i \in \mathcal{I}} \bigl( \alpha_{1,i} + 2 \alpha_2 q_i + \beta_{1,i} + 2 \beta_2 q_i \bigr)
\label{eq:SinglePairProblem}
\end{equation}

\centerline{s.t. Eq.~\eqref{eq:ScheduleConstraintMain},\eqref{eq:PowerConstraintMain},
~\eqref{eq:AnalogPower},~\eqref{eq:FrontEndConstraint}, ~\eqref{eq:CouplingConstraintMain} and~\eqref{eq:Variables1}}


The objective of the optimization problem is an affine function of $q_i$, $\alpha_i$ and $\beta_i$ for all $i \in \mathcal{I}$.
Minimum circuit power occurs if we use only one radio front end and one channel.
Since link gain of a channel does not vary across radio front ends, we can arbitrarily select a front end.
Since original system power minimization problem contains both
transmit and circuit power, it is prudent to select the channel with the best link gain.

\subsection{Use of MC-MR over NC-OFDM}

If $\alpha_2 >> \alpha_1$ and $\beta_2 >> \beta_1$, 
power consumption of ADC \& DAC dominate that of filters, mixers, etc..
Here, problem I gets reduced to the following: 
%

\begin{equation}
\min \sum_{i \in \mathcal{I}} \bigl( k_{pa} \sum_{m \in \mathcal{M}} p_i^m + 2 \alpha_2 q_i + 2 \beta_2 q_i \bigr)
\label{eq:SinglePairProblem}
\end{equation}

\centerline{s.t. Eq.~\eqref{eq:ScheduleConstraintMain},\eqref{eq:PowerConstraintMain},
~\eqref{eq:FrontEndConstraint}, ~\eqref{eq:CouplingConstraintMain} and~\eqref{eq:Variables1}}


If number of radio front ends matches the number of non-contiguous spectrum chunks, 
the optimal solution of above problem activates multiple front ends but accesses contiguous spectrum chunks in each front end.

\subsection{Use of NC-OFDM over MC-MR}

If $\alpha_2 << \alpha_1$ and $\beta_2 << \beta_1$, 
power consumption of filters, mixers, etc.. dominate that of
ADC \& DAC. In this context, problem I gets reduced to the following: 


\begin{equation}
\min \sum_{i \in \mathcal{I}} \bigl( k_{pa} \sum_{m \in \mathcal{M}} p_i^m + \alpha_{1,i}  + \beta_{1,i}  \bigr)
\label{eq:SinglePairProblem}
\end{equation}

\centerline{\eqref{eq:PowerConstraintMain}, ~\eqref{eq:AnalogPower},~\eqref{eq:FrontEndConstraint},~\eqref{eq:CouplingConstraintMain} and~\eqref{eq:Variables1}}


The optimal solution to this problem selects only one radio front end and 
accesses non-contiguous spectrum chunks through NC-OFDM to meet rate requirement.
In a practical setting, the values of $\alpha_1$, $\alpha_2$, $\beta_1$ and $\beta_2$ 
fall between the extreme cases mentioned above. In those cases,
our algorithm selects a subset of channels and assign them to different front ends to minimize system power.

\subsection{Low Complexity Algorithm}

Using the insights of Sec,~\ref{sec:Theory}, we develop a greedy algorithm here. Our algorithm can be explained simply as follows: 
\emph{Pick the channel with the highest link gain
at the first iteration. In each subsequent iteration, select exactly one channel and assign it to one front end, 
where both selection and assignment operations are the feasible operations that minimize total system power during the current
iteration}.
Table~\ref{tab:GreedyP2P} is the pseudo-code of our algorithm.
Here, $\mathcal{Y}_i \, \forall \, i \in \mathcal{I}$, is the set of channel indexes used by $i$-th front end.

\begin{table}[t]
\begin{center}
\begin{tabular}{r l} \hline
\emph{Input:} & $\mathcal{M}$, $r$, $W$, $N_0$, $g^m \, \forall m \in \mathcal{M}$ \\ \hline
\emph{Output:} & $x_i^m , p_i^m , \alpha_{1,i}, \beta_{1,i}, q_i \, , \mathcal{Y}_i \forall m \in \mathcal{M} , \forall i \in \mathcal{Y}$  \\ \hline
Line & Operation \\  \hline
1 & $ind = \arg \max_m (g^m)$ , $val = \max_m (g^m)$. \\ \hline
2 & $x_1^{ind} = 1$, $\mathcal{Y}_1 = \{ind\}$, $q_1 = W$, $\alpha_{1,1} = \alpha_1$, $\beta_{1,1} = \beta_1$ \\ \hline
3 & $P_{tot} = k_{pa} \bigl(2^{\frac{r}{W}} - 1 \bigr) \frac{N_0}{val} + \alpha_1 + 2 \alpha_2 W + \beta_1 + 2 \beta_2 W$ \\   \hline
4 & while (\emph{true})  \\   \hline
5 & \hspace{3 mm} $\forall m \in \mathcal{M}$ \\ \hline
6 & \hspace{6 mm} $\forall i \in \mathcal{I}$ \\   \hline
7 & \hspace{9 mm} $\widetilde{\alpha}_{1,i} = \alpha_1$, $\widetilde{\beta}_{1,i} = \beta_i$ , $\tilde{q}_i = q_i$, $\tilde{x}_i^m = x_i^m$ \\  \hline
8 & \hspace{9 mm} if $ m \notin \mathcal{Y}_i \, \forall \, i \, \in \, \mathcal{I}$  \\  \hline
9 & \hspace{12 mm}  $MaxInd = max\{\mathcal{Y}_i,m\}$ , $MinInd = min\{\mathcal{Y}_i,m\}$ \\  \hline
10 & \hspace{12 mm}   $\tilde{q}_i = W (MaxInd - MinInd + 1)$ , $\tilde{x}_i^m = 1$ \\  \hline
11 & \hspace{12 mm} $\widetilde{\alpha}_{1,i} = \alpha_1$ , $\widetilde{\beta}_{1,i} = \beta_1$. \\ \hline
12 & \hspace{12 mm}  $\tilde{f}_j^n = \frac{r}{\sum_{j \in \mathcal{I}} \sum_{n \in \mathcal{M}} \tilde{x}_j^n} \tilde{x}_j^n \, \, \forall j \in \mathcal{I} , \, \forall n \in \mathcal{M}$ \\  \hline
13 & \hspace{12 mm} $\tilde{p}_j^n = (2^{\frac{\tilde{f}_j^n}{W}}-1) \frac{N_0 W}{g^n} \, \, \, \forall j \in \mathcal{I} \, , \, \forall n \in \mathcal{M}$ \\ \hline
14 & \hspace{12 mm}  $\tilde{P}_i^m = \sum_{j \in \mathcal{I}} \bigl(\alpha_{1,j} + \alpha_2 q_j + \beta_{1,j} + \beta_2 q_j + $ \\
& \hspace{12 mm} $k_{pa} \sum_{n \in \mathcal{M}} \tilde{x}_j^n \bigr) $ \\ \hline
15 & \hspace{3 mm} $[minP, minC, minI] = \min_m \min_i ( \tilde{P}_i^m)$   \\ \hline
16 & \hspace{3 mm} if $minP \leq P_{tot}$, \\ \hline
17 & \hspace{6 mm} $P_{tot} = minP$, $x_{minI}^{minC} = 1$, $\alpha_{1,minI} = \alpha_1$, $\beta_{1,minI} = \beta_1$. \\ \hline
18   & \hspace{6 mm} $\mathcal{Y}_{minI} = \{ \mathcal{Y}_{minI}, minC \}$. \\ \hline
19   & \hspace{6 mm} $MaxInd = max \{\mathcal{Y}_{minI} \}$. $MinInd = min \{ \mathcal{Y}_{minI} \}$.  \\ \hline
20 & \hspace{6 mm} $q_i = W (MaxInd - MinInd + 1)$. \\ \hline
21 & \hspace{3 mm} else \\ \hline
22 & \hspace{6 mm} \emph{break}  \\ \hline
23  &  $\min \sum_{i \in \mathcal{I}} \sum_{m \in \mathcal{M}} p_i^m$ \\ \hline
24 & \hspace{3 mm} s.t. $\sum_{i \in \mathcal{I}} \sum_{m \in \mathcal{M}} W log(1 + \frac{p_i^m g^m}{N_0 W}) \geq r$, \\ \hline
25 & \hspace{3 mm} $p_i^m \leq A x_i^m \, , \, p_i^m \, \geq \, 0 \, \forall \, m \in \mathcal{M} \, , \, \forall i \in \mathcal{I}$.  
\end{tabular}
\end{center}
\caption{Greedy Algorithm to Minimize System Power in a Multi-front end radio enabled 
point-to-point link}
\label{tab:GreedyP2P}
\end{table}

\begin{figure*}[t]
\begin{centering}
\includegraphics[scale=0.5]{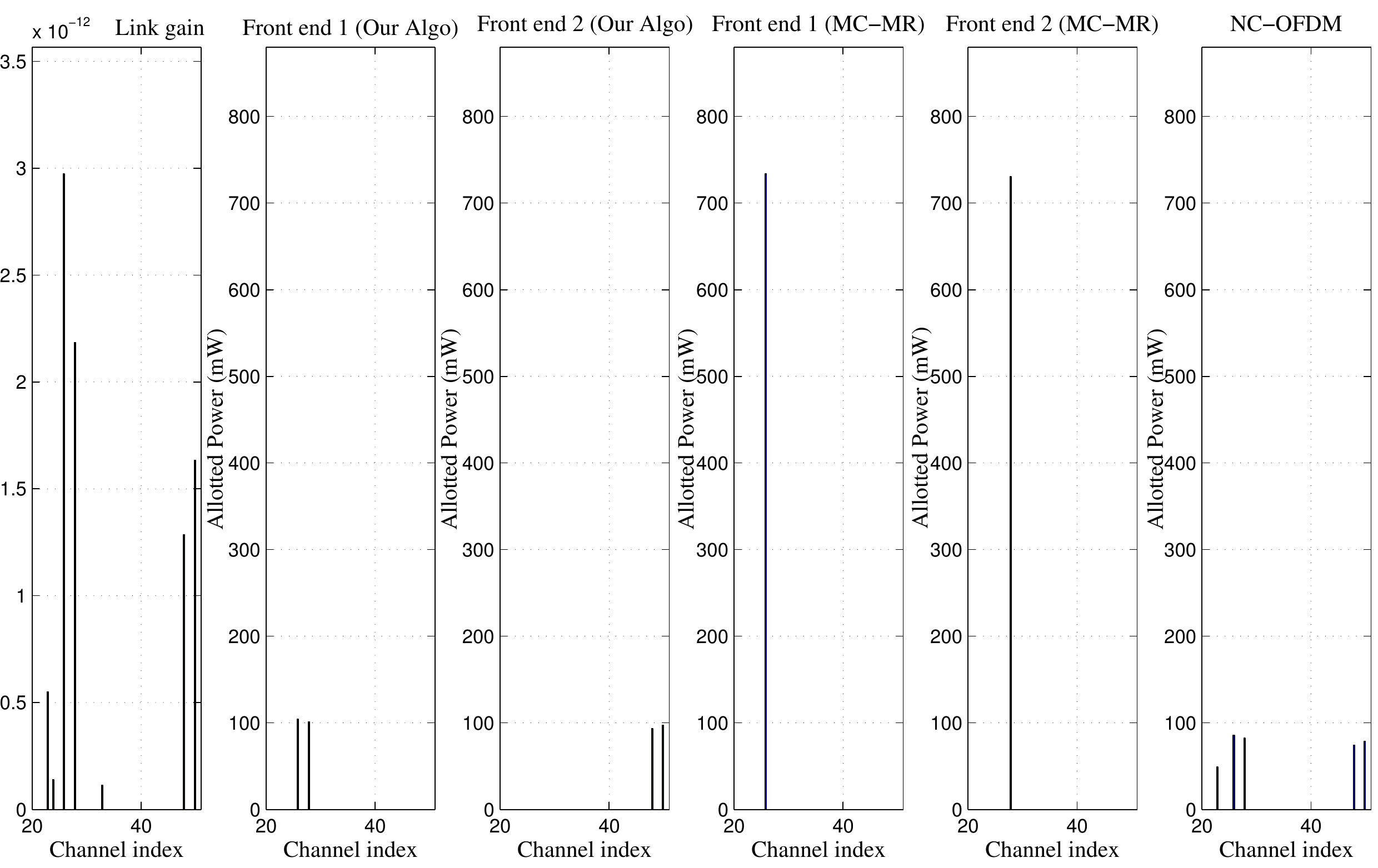}
\caption{Power allocation among TV white space channels for portable devices of Cambridge~\cite{ShowWhiteSpace} (Demand = 75 Mbps)}
\label{fig:Cambridge}
\end{centering}
\end{figure*}

At first, we select the channel with the best link gain (line $1$). Without loss of generality,
we assign this channel to first front end (line $2$) and find associated total power (line $3$).
Thereafter, an infinite while loop starts (line $5$) where each loop
looks at all possible combinations of channel and front end assignment (line $6$-$7$)
and selects the channel-front end combination that minimizes system power during current iteration.
For each channel-front end combination, we calculate spectrum span (line $9$-$10$), 
flow allocation (line $12$), power allocation (line $13$) and total power (line $14$). 
Line $15$ compares total power values among all channel-front end combinations
and finds the channel and front end combination that leads to minimum at the current iteration.
If the current value of minimum total power is less than our previously stored value (line $16$),
we update total power, scheduling and spectrum span to accommodate current assignment (line $17$-$20$).
Otherwise, infinite loop breaks (line $21$-$22$). 
The infinite while loop of line $4$-$22$ provides scheduling variables, spectrum span and circuit power
for all channels and front ends. Line $23$-$25$ allocates optimal power across these scheduled channels and front ends.
We assume equal flow per channel (line $12$)
during the inside iterations of while loop  to reduce complexity but our final solution contains optimal flow
and power allocation across channels and front ends (line $23$-$25$).

\subsubsection{Computational Complexity}

Overall complexity comes from the while loop of line $5$-$22$ and the convex minimization program of line $23$-$25$. 
The loops of line $6$ and $7$ run $M$ and $I$ times respectively. For each channel-front end combination,
we find power and flow values for all activated channels $(O(M))$. The outer while loop of line
$5$ runs at most $M$ times since each iteration will either select a better channel or the loop will break.
Hence, the while loop of line $5$-$22$ runs at most $(O(M^3 I))$ times.
The convex optimization program of line $23$-$25$ contains $M I$ number of constraints and can be solved in
$(O (MI \log (MI)))$ time~\cite{Boyd}. Hence, overall complexity is $(O(M^3 I))$.

%

\subsection{Other Algorithms}

We compare our algorithm with two other algorithms. First algorithm
only focuses on MC-MR platforms and accesses contiguous spectrum chunks in each
of its radio front end~\cite{MC-MR1}. This algorithm puts enough power in contiguous spectrum chunks of each front end
to meet rate requirement. We term this approach 'MC-MR'. Second algorithm only focuses on NC-OFDM technology,
spreads power across all ``good" channels and
accesses multiple non-contiguous spectrum chunks with one front end ~\cite{Nazmul_NCOFDM1}.
We term this approach 'NC-OFDM'. Both these approaches try to minimize the transmitted power at radio frequency. 
We compare the performance of both these algorithms with our algorithm in next section.

\section{Simulation Results}  \label{sec:Simulation}

We focus on a point-to-point link where both transmitter and receiver are portable devices and they access 
white space channels to meet rate requirement. Table~\ref{tab:ChannelList}
shows the indexes and center frequencies of available channels~\cite{ShowWhiteSpace}. Here, 
$\mathcal{M} = \{ 23, 24, 26, 28, 33, 48, 50 \}$, $M = 7$. We assume that both transmitter and receiver
are equipped with two front ends, i.e., $\mathcal{I} = \{1, 2 \}$ and $I = 2$.
Table~\ref{tab:ParameterValues} shows the values of different parameters that we used in this section.
A detailed explanation of these values can be found in~\cite{Nazmul_NCOFDM1}.

\begin{table}[t]
\begin{center}
\begin{tabular}{|l|l|l|l|l|l|l|l|} \hline
Channel Index      & $23$ & $24$ & $26$ & $28$ & $33$ & $48$ & $50$ \\ \hline
Center Freq. (MHz) & $527$ & $533$ & $545$ & $557$ & $587$ & $677$ & $689$ \\ \hline
\end{tabular}
\end{center}
\caption{Available TV channels for portable devices in Cambridge, MA~\cite{ShowWhiteSpace} }
\label{tab:ChannelList}
\end{table}

\begin{table}[t]
\begin{center}
\begin{tabular}{|l|l|l|l|l|l|l|} \hline
Parameters & $\alpha_1$ & $\alpha_2$ & $\beta_1$ & $\beta_2$ & $k_{pa}$ & $W$   \\ \hline
Values &  $45.4$ & $7.2$ & $282.3$ & $5.5$ & $10.67$ & $6$  \\ 
& mW & mW/MSPS & mW & mW/MSPS & & MHz \\ \hline
\end{tabular} 
\end{center}
\caption{Values of different parameters used in Sec.~\ref{sec:Simulation}}
\label{tab:ParameterValues}
\end{table}

The left subfigure of Fig.~\ref{fig:Cambridge} shows the link gain across these channels.
We assumed $500$m distance between two nodes, path loss with exponent $3$ and $15$ dB random variation
to generate link gain across these channels. We assume $75$ Mbps demand ($r$) for this simulation scenario. The 2nd and 3rd subfigures (from left) of Fig.~\ref{fig:Cambridge}
show power allocation variables of our algorithm across both front ends. Our algorithm selects ``better"  
channels - in terms of link gain - among the whole list, accesses ``nearby" non-contiguous spectrum chunks
in each front end and allocates transmit power to meet demand requirement.

`MC-MR' algorithm can only access contiguous spectrum chunks in each of its front end.
Since channel $23$ \& $24$ - the only contiguous chunk with two channels - have poor link gain, this algorithm selects 
channel $26$ and $28$ in each of its radio front end. Since demand is much higher than used bandwidth,
The 4th and 5th subfigures (from left) of Fig.~\ref{fig:Cambridge} shows that MC-MR algorithm allocates very high amount of power in each channel and meets rate requirement. 
The rightmost sub-figure of Fig.~\ref{fig:Cambridge} shows that 
`NC-OFDM' algorithm spreads power across all 'good' channels, accesses multiple non-contiguous spectrum chunks \emph{with one front end}
and spans a wide amount of spectrum. As a result, `NC-OFDM' algorithm consumes high power in ADC/DAC circuits~\cite{Nazmul_NCOFDM1}.

\begin{figure}[t]
\begin{centering}
\includegraphics[scale=0.35]{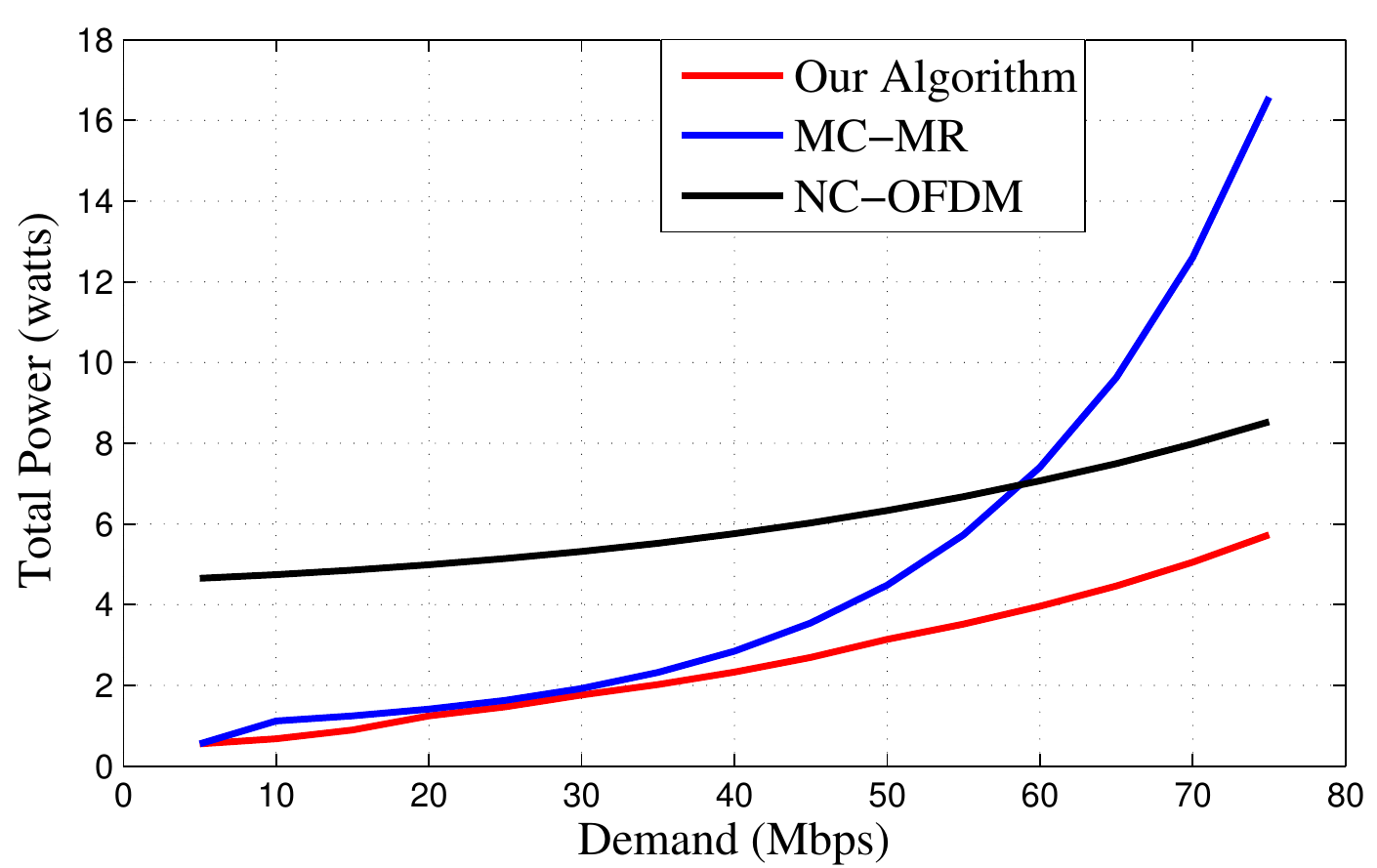}
\caption{Comparison of system power consumption among different approaches}
\label{fig:PowerComparison}
\end{centering}
\end{figure}

Fig.~\ref{fig:PowerComparison} compares the total power consumption of our algorithm with that
of `MC-MR' and 'NC-OFDM' algorithm for different demand values. As evident from Fig.~\ref{fig:Cambridge},
`MC-MR' algorithm consumes a high amount of transmit power for high demand.
When demand is low (e.g. $5$ Mbps), both our algorithm and `MC-MR' approach access only one channel using one front end
and consume equal system power. `NC-OFDM' approach spans wider amount of spectrum and, hence, consumes excessive circuit
power in ADC and DAC. Our algorithm reaps the inherent benefits of `MC-MR' and `NC-OFDM' approach and combines
them to minimize system power across all scenarios.

\section{Conclusion} \label{sec:Conclusion}

MC-MR and NC-OFDM allow nodes to access non-contiguous spectrum chunks but increase circuit power
either by activating multiple radio front ends or by spanning wider spectrum.
This work performed joint optimal power control and scheduling to minimize the system power of a point-to-point
link where both nodes are equipped with MC-MR and NC-OFDM technology.
We designed a low complexity greedy algorithm $(O(M^3 I))$ where $M$ and $I$ denote the number of channels and radio front ends respectively.

The number of available channels and non-contiguous 
spectrum chunks vary across different places of USA~\cite{ShowWhiteSpace}. 
Our work can be extended to investigate how a radio with limited number of front ends
performs in non-contiguous spectrum access across different US cities.



\bibliographystyle{IEEEbib}
\bibliography{BibNCOFDMA}

\end{document}